\def\spose#1{\hbox to 0pt{#1\hss}}

\def\multleft#1{\hbox to size{\vbox {\halign {\lft{##}\cr #1}}\hfill}\par}
\def\multright#1{\hbox to size{\vbox {\halign {\rt{##}\cr #1}}\hfill}\par}

\def\today{\ifcase\month\or January\or February\or March\or April\or May\or
      June\or July\or August\or September\or October\or November\or December\fi
      \space\number\day, \number\year}
\def\s{\hbox{\phantom{5}}}	

\def\cm{{\rm\thinspace cm}}

\def\erg{{\rm\thinspace erg}}

\def\keV{{\rm\thinspace keV}}
\def\km{{\rm\thinspace km}}
\def\kpc{{\rm\thinspace kpc}}

\def\Mpc{{\rm\thinspace Mpc}}
\def\Msun{\hbox{$\rm\thinspace M_{\odot}$}}
\def\pc{{\rm\thinspace pc}}

\def\s{{\rm\thinspace s}}

\def\yr{{\rm\thinspace yr}}
\def\Myr{{\rm\thinspace Myr}}
\def\Gyr{{\rm\thinspace Gyr}}

\def\ergpcmsqps{\hbox{$\erg\cm^{-2}\s^{-1}\,$}}

\def\ergps{\hbox{$\erg\s^{-1}\,$}}

\def\kmps{\hbox{$\km\s^{-1}\,$}}

\def\Msunpyr{\hbox{$\Msun\yr^{-1}\,$}}

\def\kmpspMpc{\hbox{$\kmps\Mpc^{-1}$}}

\def\H2{\hbox{H$_{2}$}}

\def\Pa{Pa$\alpha$}
\def\Ha{H$\alpha$}
\def\Bg{Br$\gamma$}
\documentclass[usegraphicx]{mn2e}

\usepackage{times}
\usepackage{amssymb}
\usepackage{eso-pic}
\include{defn}

\begin{document}
\hsize=6truein
\title[Molecular accretion in the core of the galaxy cluster 2A 0335+096]
{Molecular accretion in the core of the galaxy cluster 2A 0335+096}
\author[R.J.~Wilman et al.]
{\parbox[]{6.in} {R.J.~Wilman$^{1,2}$, A.C.~Edge$^{3}$, P.J. McGregor$^{4}$, B.R.~McNamara$^{5}$. \\ \\
\footnotesize
1. School of Physics, University of Melbourne, Parkville, Victoria 3010, Australia \\ 
2. Centre for Astrophysics \& Supercomputing, Swinburne University of Technology, Hawthorn, Victoria 3122, Australia \\
3. Department of Physics, University of Durham, South Rd, Durham, DH1 3LE \\
4. Research School of Astronomy \& Astrophysics, The Australian National University, Cotter Rd, Weston Creek, ACT 2611, Australia
5. Department of Physics \& Astronomy, University of Waterloo, 200 University Avenue West, Ontario, N2L 3G1, Canada \\}}

\maketitle

\begin{abstract}
We present adaptive optics-assisted K-band integral field spectroscopy of the central cluster galaxy in 2A 0335+096 (z=0.0349). 
The \H2~v=1--0~S(1) emission is concentrated in two peaks within 600\pc~of the nucleus and fainter 
but kinematically-active emission extends towards the nucleus. The \H2~is in a rotating structure which aligns with, 
and appears to have been accreted from, a stream of \Ha~emission extending over 14\kpc~towards a companion galaxy. 
The projected rotation axis aligns with the 5~GHz radio lobes.

This \H2~traces the known $1.2\times 10^{9}$\Msun~CO-emitting reservoir; limits on the \Bg~emission confirm that the \H2~emission 
is not excited by star formation, which occurs at a rate of less than 1\Msunpyr~in this gas. If its accretion onto the black hole 
can be regulated whilst star formation remains suppressed, the reservoir could last for at least 1\Gyr; the simultaneous 
accretion of just $\sim 5$~per cent of the gas could drive a series of AGN outbursts which offset X-ray cooling in the cluster 
core for the full $\sim 1$\Gyr. Alternatively, if the regulation is ineffective and the bulk of the \H2~accretes within a few orbital 
periods (25--100\Myr), the resulting $10^{62}$\erg~outburst would be among the most powerful cluster AGN outbursts 
known. In either case, these observations further support cold feedback scenarios for AGN heating. 
\end{abstract}


\section{INTRODUCTION}
It is now widely accepted that `feedback' from radio-loud active galactic nuclei (AGN) keeps the bulk of 
the diffuse X-ray gas hot in the cores of clusters of galaxies (e.g. Rafferty et al.~2008; Cavagnolo et al.~2008).
Although the underlying plasma microphysics are complex (McNamara \& Nulsen~2007), it is 
also clear that AGN heating is not always and everywhere 100 per cent effective. Some gas does succeed in cooling, perhaps in an episodic way when the AGN is in quiescence, and accumulates in a reservoir of cool ionized and molecular gas (e.g. Edge~2001; Edge et al.~2002; Johnstone et al.~2007), giving rise to star formation (Crawford et al.~1999; O'Dea et al.~2008). The associated emission line nebulosity has been the subject of many investigations over the years, focussing variously on the kinematics of the gas, the origin of its filamentary morphology, its relationship to the radio source, the role of galaxy interactions and mergers, 
and the nature of the excitation source. Progress on the latter front came with the simulations of 
Ferland et al.~(2009), who showed that extra heating by dissipative MHD processes or non-thermal excitation by high energy cosmic 
rays could account for the unusually strong ro-vibrational \H2~lines and other emission line ratios which are at odds with the
predictions of photoionization models.

An issue that has seldom been addressed is whether this cool gas reservoir plays an active role
in the feedback cycle or whether it is just an interesting byproduct thereof, serving merely to show that cooling and AGN feedback are taking place but of no greater significance. It is of particular interest to know whether the cool gas actually fuels the accretion on to the supermassive black hole. For many years, it has been hypothesized that `hot feedback' in the form of quasi-spherical Bondi accretion of partially-cooled X-ray gas could be the mechanism at work (e.g. Allen et al.~2006;  Narayan \& Fabian~2011). This model has, however, encountered problems and appears to be energetically infeasible except in low-power systems (e.g. Soker~2010; Rafferty et al.~2006; McNamara et al.~2011), leading 
to the development of the `cold feedback' model in which the AGN accretes directly from the cool gas reservoir (see Pizzolato \& Soker~2005,~2010 and references therein). Non-linear overdensities in the X-ray gas throughout the inner 30\kpc~are assumed to cool and flow steadily inwards. 
Blobs with an overdensity exceeding a critical factor $\simeq 2$ relative to the ambient medium undergo catastrophic cooling, contract and enter a free-fall plunge over the final 10\kpc, creating a reservoir of cool gas for star formation and black hole accretion. Pizzolato \& Soker~(2010) demonstrated that the infalling blobs efficiently lose angular momentum via drag forces and mutual collisions, thereby removing a potential obstacle to the accretion. The means by which the molecular gas couples to the AGN output -- and thereby regulates the cold feedback process -- are, however, unclear; outflows driven by jets or collimated winds may be involved (e.g. Soker \& Pizzolato~2005; Wagner \& Bicknell~2011).

Aspects of the cold feedback model are steadily gaining observational support. In NGC 1275, Lim, YiPing \& Dinh-V-Trung~(2008) observed filaments of CO emission extending out to radii of 8\kpc, coincident with the coolest X-ray gas either side of the radio cavities. This cool molecular gas is in gravitational free-fall and appears to feed a disk of hot \H2~at radii $<50$\pc~(Wilman et al.~2005; McGregor et al.~2007), thereby closing the feedback loop. In a study with near-infrared integral field spectroscopy, Wilman et al.~(2009) observed three well-known local cooling flow clusters: A1664, A2204 and PKS 0745-191. The findings were consistent with the three systems being captured at distinct phases of the cold feedback cycle.

Building on these studies, we now present K-band integral field spectroscopy (IFS) of the cental cluster galaxy in the nearby cluster 2A 0335+096, 
focussing on the \H2~line emission. The observations were conducted with the Near-Infrared Integral Field Spectrograph (NIFS) on 
the {\em Gemini North} telescope, used in conjunction with the ALTAIR adaptive optics facility to maximise spatial resolution in the circumnuclear regions of this low redshift system (z=0.0349). Optical IFS by Hatch et al.~(2007) showed that the \Ha~emission arises in a bar with twin peaks 
within 1\arcsec~of the nucleus of the brightest cluster galaxy (BCG). This bar is part of a diffuse stream of \Ha~extending over 20\arcsec~(14\kpc), to and 
beyond a companion cluster elliptical galaxy 4\kpc~away in projection, for which Donahue et al.~(2007) presented narrow-band \Ha~imaging and long-slit 
spectroscopy and archival ultraviolet imaging, building on earlier \Ha~and multi-colour imaging by Romanishin \& Hintzen~(1988). They confirmed that the redshifts of the BCG and its companion lie within 100\kmps~of each other and that the pair are probably 
interacting. The estimated timescale for the interaction is comparable with the age of the radio source, suggesting that the latter was triggered by the interaction. Both galaxies exhibit CO emission with interferometry by Edge \& Frayer~(2003) revealing that the bulk of it, $1.2 \times 10^{9}$\Msun, is situated within the central 3.5\kpc~of the BCG. The latest {\em Chandra} X-ray and radio observations (Sanders, Fabian \& Taylor~2009) showed that the X-ray emission from the coolest 0.5\keV~gas exhibits a 
close spatial correspondence with the \Ha~stream, with the AGN radio emission emerging perpendicular to it. Mid-infrared photometry and spectroscopy with {\em Spitzer} have yielded measurements of the star formation rate in this system of 2.1\Msunpyr~(O'Dea et al.~2008) and 0.7\Msunpyr~(Donahue et al.~2011), respectively.

\section{OBSERVATIONS AND DATA REDUCTION}
Observations of 2A 0335+096 with {\em Gemini}-NIFS were performed in queue mode on 2009 October 10, 11 and 13 yielding a total of 17 useable on-source exposures each of 10 minutes duration. Sky subtraction was enabled via an object-object-sky exposure nodding pattern using a sky region 30\arcsec~from the BCG. Four additional on-source observations for the programme were obtained on 2009 September 8 and October 12 but were not included in our analysis due to technical problems during the observations. The ALTAIR adaptive optics (AO) unit was used with a laser guide star, operating under natural seeing in the range 0.2--0.5\arcsec. 

NIFS operates on the image slicing principle with a $3 \times 3$\arcsec~field of view sampled by 0.103\arcsec-wide slitlets and a pixel scale of 0.043\arcsec~along the slits. A position angle of 335 degrees was used to align the slits with the major axis of the BCG. The K-band grating was used, spanning 1.99--2.40\micron~at a dispersion of $2.1338 \times 10^{-4}$\micron~per pixel, corresponding to a velocity resolution of 60\kmps. 

The data were processed in IRAF using the Gemini and NIFS-specific routines. The task NFTELLURIC was used to correct for the atmospheric
absorption via a standard star, and the task NIFCUBE was used to generate a single datacube for each of the on-source observations with the 
default re-sampling to 0.05\arcsec~square pixels. The AO-corrected point spread function (PSF) was measured using the telluric standard stars and in all cases was found to be sharply peaked with a full width at half maximum (FWHM) of 0.11\arcsec~along the slitlets; in the perpendicular direction the 0.103\arcsec~slit-width undersamples the diffraction limit and the PSF is consequently less sharply peaked, with a FWHM of approximately 0.19\arcsec. The encircled energy fractions within radii of 0.1 and 0.2\arcsec~were 43 and 63 per cent, respectively. The individual on-source science frames were median combined to produce a single datacube. The latter was re-sampled to 0.1\arcsec~pixels for compatibility with the PSF and enhanced signal-to-noise. The spectra were not flux calibrated. IDL was used for all subsequent analysis. During the latter, residuals from sky emission line subtraction were found across the 
entire field of view in two adjacent spectral pixels at 2.191\micron, within the profile of an \H2~emission line in 2A 0335+096. 
The feature was present in all the individual datacubes and was mitigated by interpolating across pixels either side of the glitch.

In a cosmology with $H_{\rm{0}}=70$\kmpspMpc, $\Omega_{\rm{M}}=0.3$ and $\Omega_{\rm{\Lambda}}=0.7$ the
spatial scale in 2A 0335+096 (z=0.0349) is 700\pc~per arcsec.

\section{RESULTS}

\subsection{Morphology \& kinematics of \H2~v=1-0~S(1) emission}
The only detectable emission lines in the NIFS spectra are part of the v=1-0~S ro-vibrational series of \H2, namely 
S(1), S(2) and S(3). The S(3) line is redshifted into an atmospheric absorption band so our analysis focusses on
the remaining ortho transition, S(1) at 2.1218\micron~(rest) which is redshifted to a clean region of transmission at 
2.195\micron. 

\begin{figure*}
\begin{center}
{
\includegraphics[width=7.5cm,angle=0]{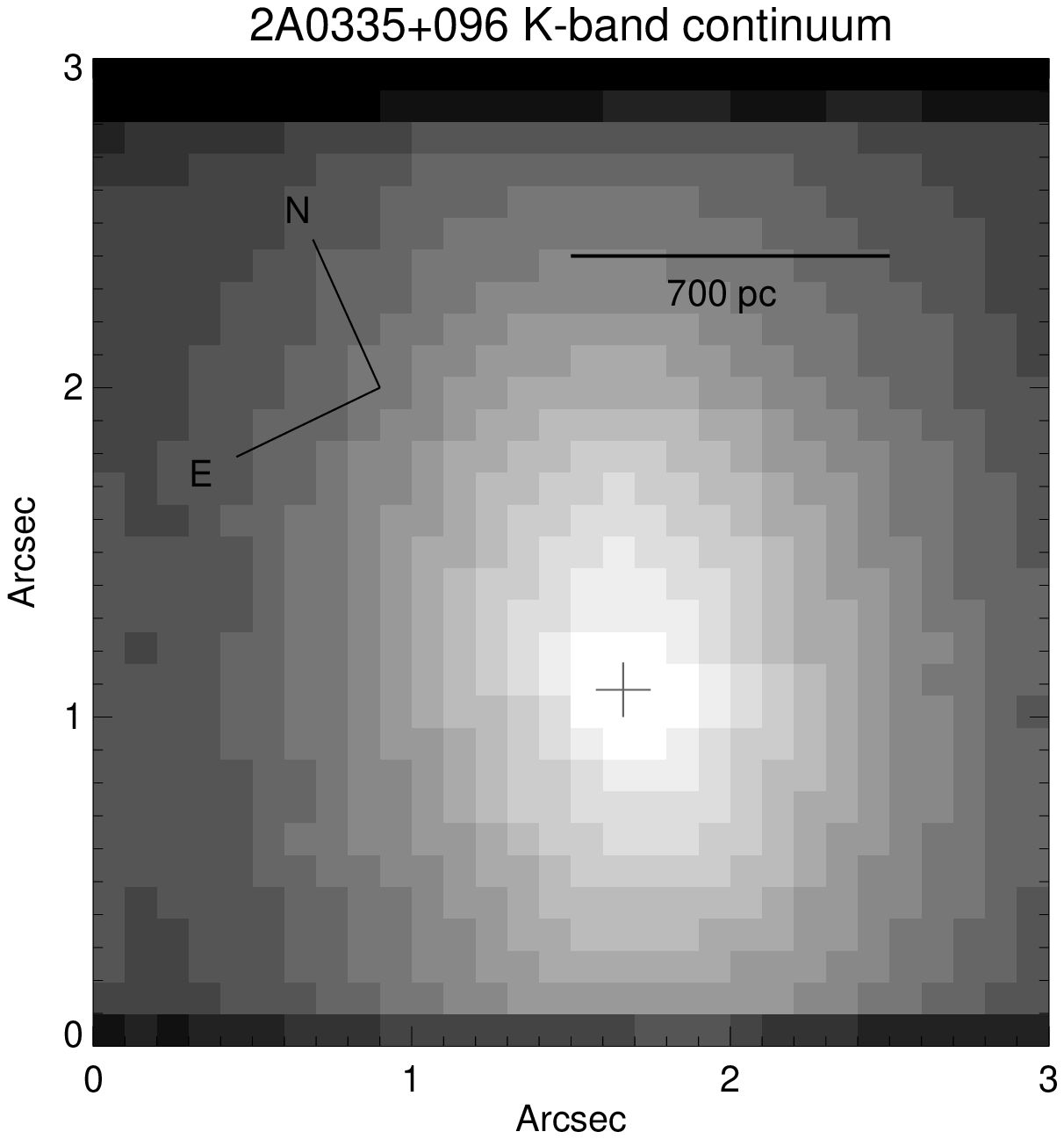}
\includegraphics[width=7.5cm,angle=0]{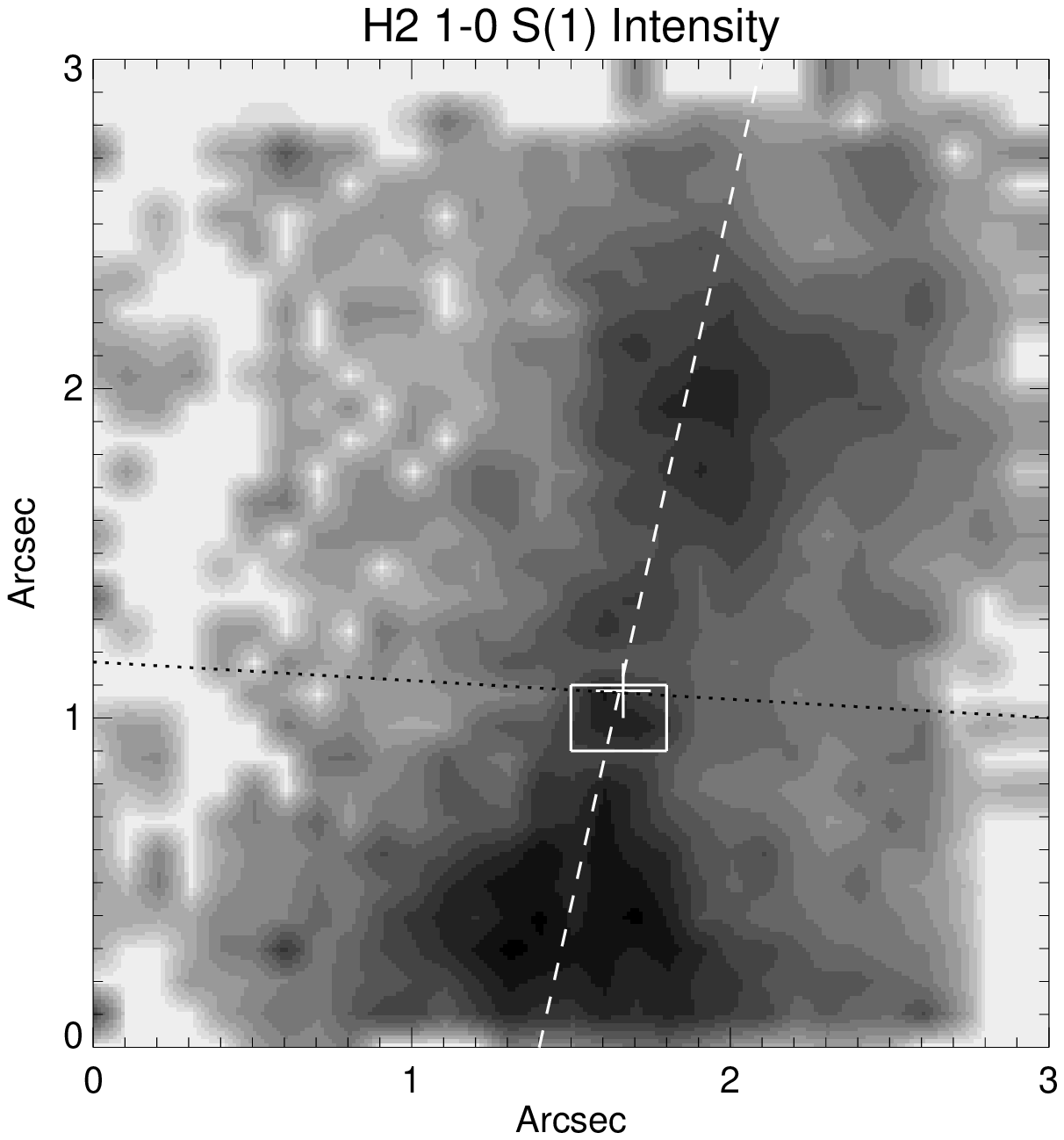}
}
\end{center}
\vspace*{0.3cm}
\begin{center}
{
\includegraphics[width=7.5cm,angle=0]{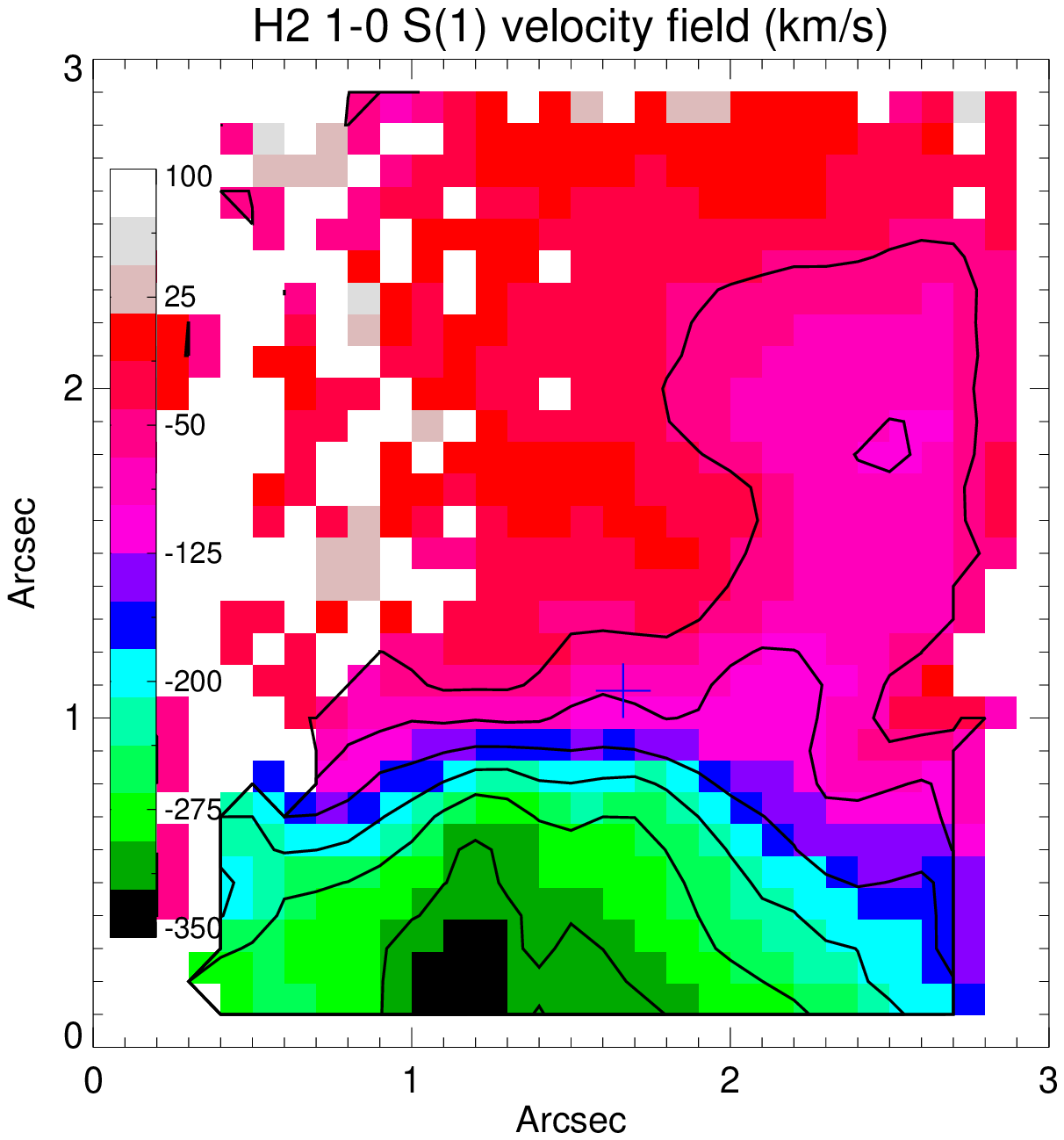}
\includegraphics[width=7.5cm,angle=0]{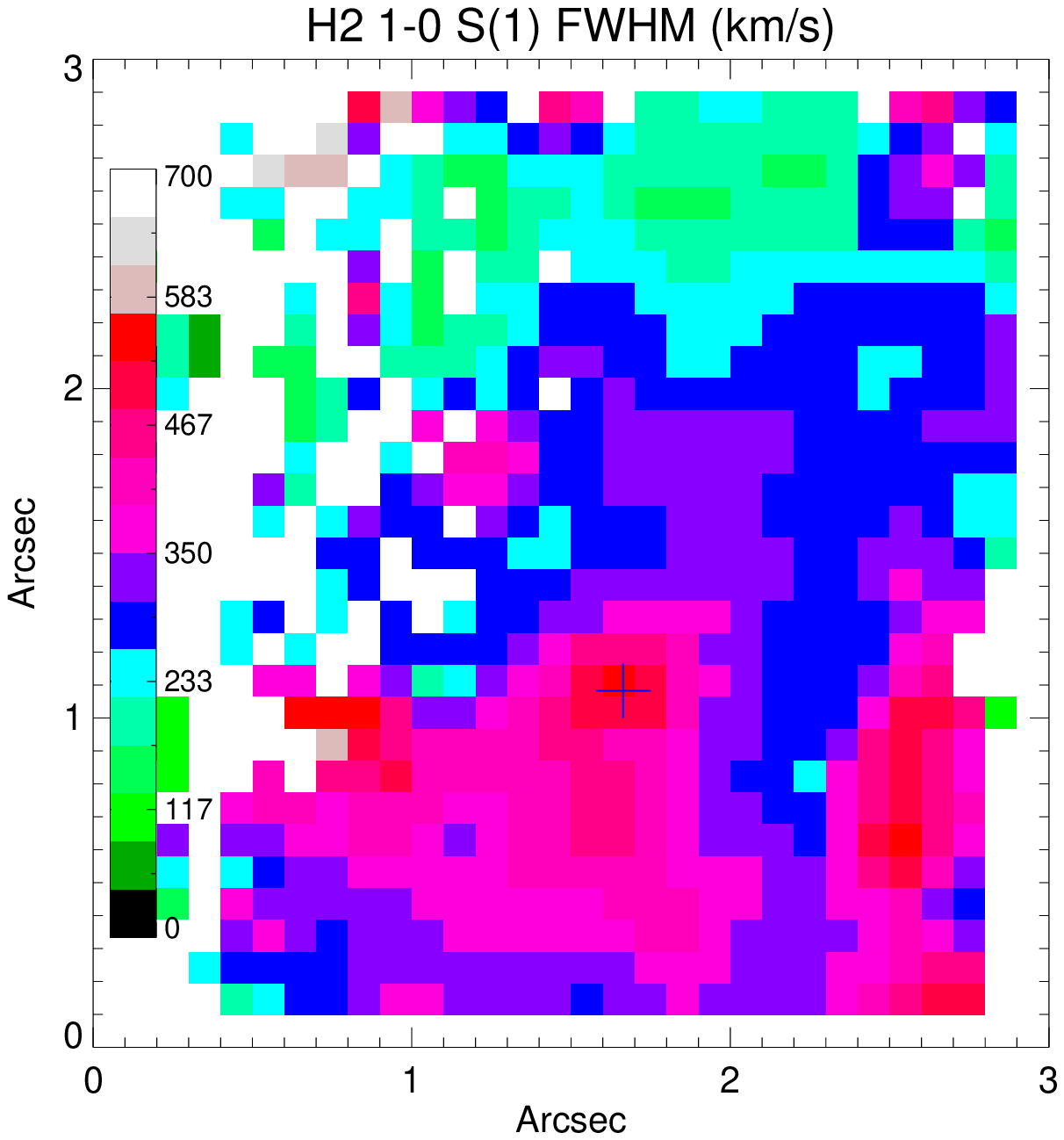}
}
\end{center}
\caption{Top row: reconstructed K-band continuum image of 2A 0335+06 from the NIFS datacube-- the cross marks the nucleus; 
\H2~v=1-0~S(1) emission line image (resampled to 0.0125\arcsec~pixels to remove visual pixellation; the white dashed line indicates the \H2~major axis
used for the position-velocity plot in Fig.~\ref{fig:2AH2kinemcut}), and the black dotted line indicates the position angle of the 5~GHz radio emission lobes in Sanders et al.~(2009);  the white box indicates the nuclear region for which a spectrum is shown in Fig.~\ref{fig:2AH2spec}. Lower row: \H2~velocity field relative to an assumed zeropoint of z=0.0349; \H2~velocity dispersion field (FWHM). Both kinematic plots have been smoothed by a flux-weighted
average of the surrounding 8 spaxels. Isovelocity contours overlaid on the \H2~velocity field span --350 to --50\kmps~at 50\kmps~intervals.}
\label{fig:2AH2}
\end{figure*}

\begin{figure}
\begin{centering}
\includegraphics[width=0.5\textwidth,angle=0]{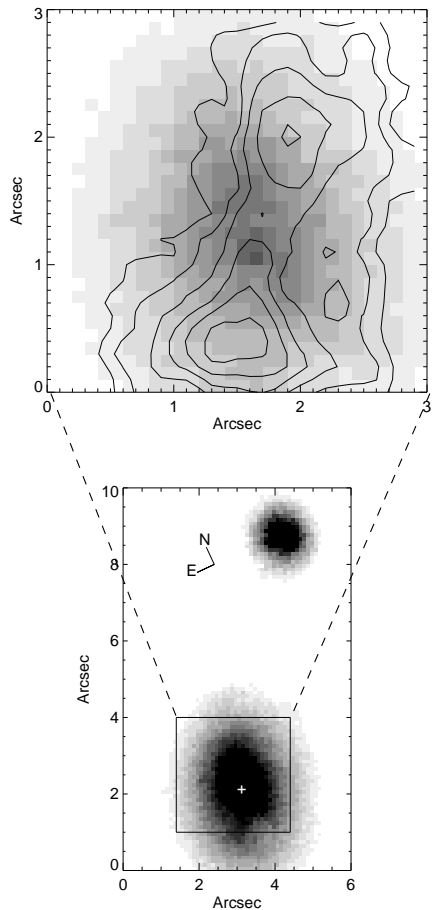}
\caption{Top: contours of the \H2~emission in 2A 0335+096 overlaid on the {\em HST} F606W image; Lower: a wide-field view of the same HST image with the NIFS field-of-view overlaid and the greyscale adjusted to show the dust absorption feature 1-1.5\arcsec~south of the nucleus and the companion cluster galaxy.}
\label{fig:2AHST}
\end{centering}
\end{figure}

The results derived from the automated fitting of a single gaussian to the \H2~v=1-0~S(1) emission line in the individual
spaxels are shown in Fig.~\ref{fig:2AH2}. The \H2~comprises two prominent clumps straddling the nucleus some 600\pc~to the north-west
and south-east, the latter perhaps splitting into two sub-clumps. Weaker emission extends from both clumps towards the nucleus and 
diffuse emission envelops the entire structure, particularly on the south-western side. In Fig.~\ref{fig:2AHST} we overlay contours of 
the \H2~emission on an {\em Hubble Space Telescope} WFPC2 F606W image of 2A 0335+096. This highlights a region of dust absorption 
1--1.5\arcsec~south of the nucleus at the lower edge of the NIFS field and the companion elliptical galaxy 4\kpc~away in projection. 
The \H2~emission is oriented roughly perpendicular on the plane of the sky to the weak lobes of 5~GHz radio emission (Sanders et al.~2009) 
which extend for 12\arcsec~(8.4\kpc) either side of the nucleus.

\begin{figure}
\begin{centering}
\includegraphics[width=6.5cm,angle=0]{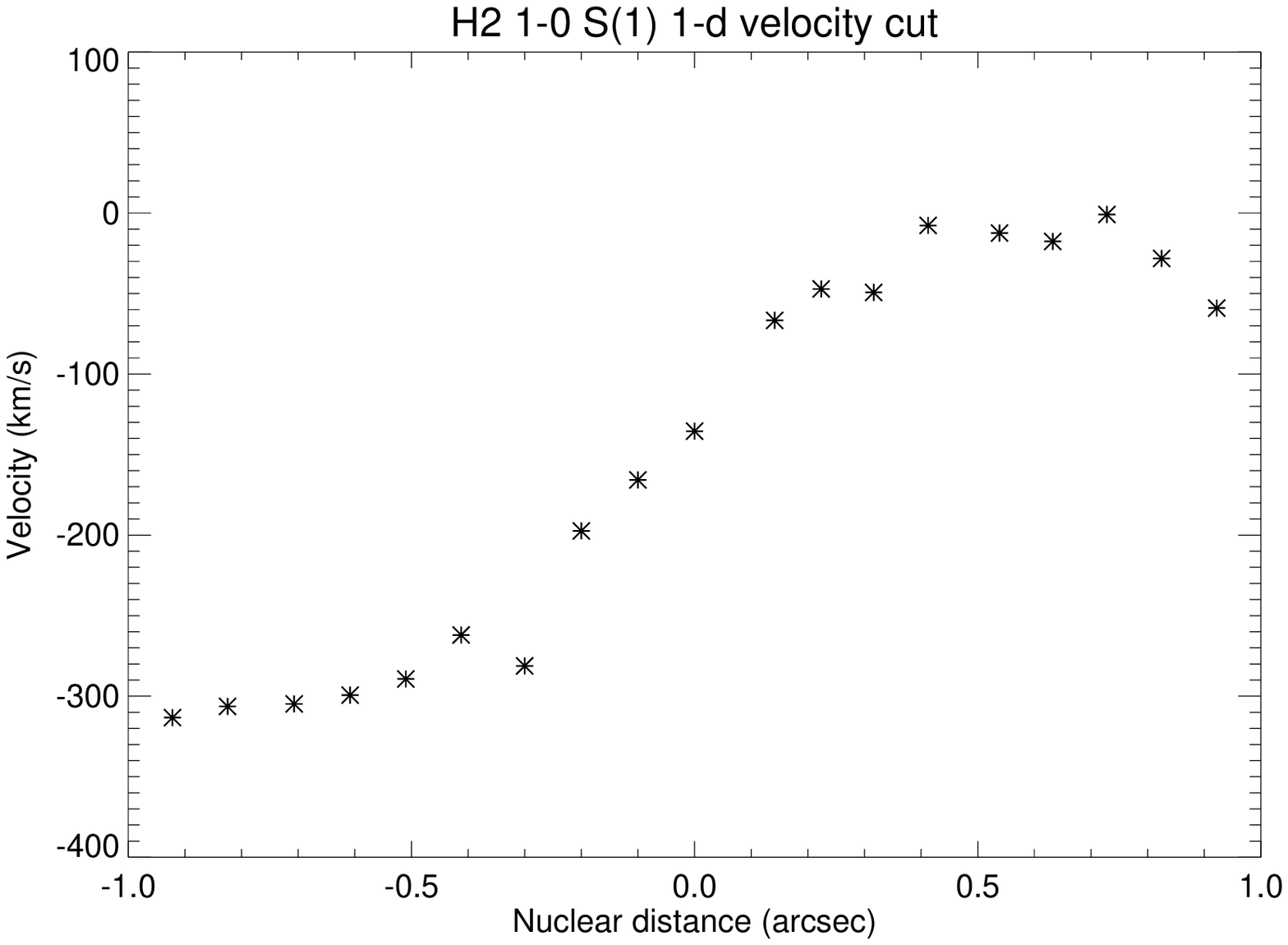}
\includegraphics[width=6.5cm,angle=0]{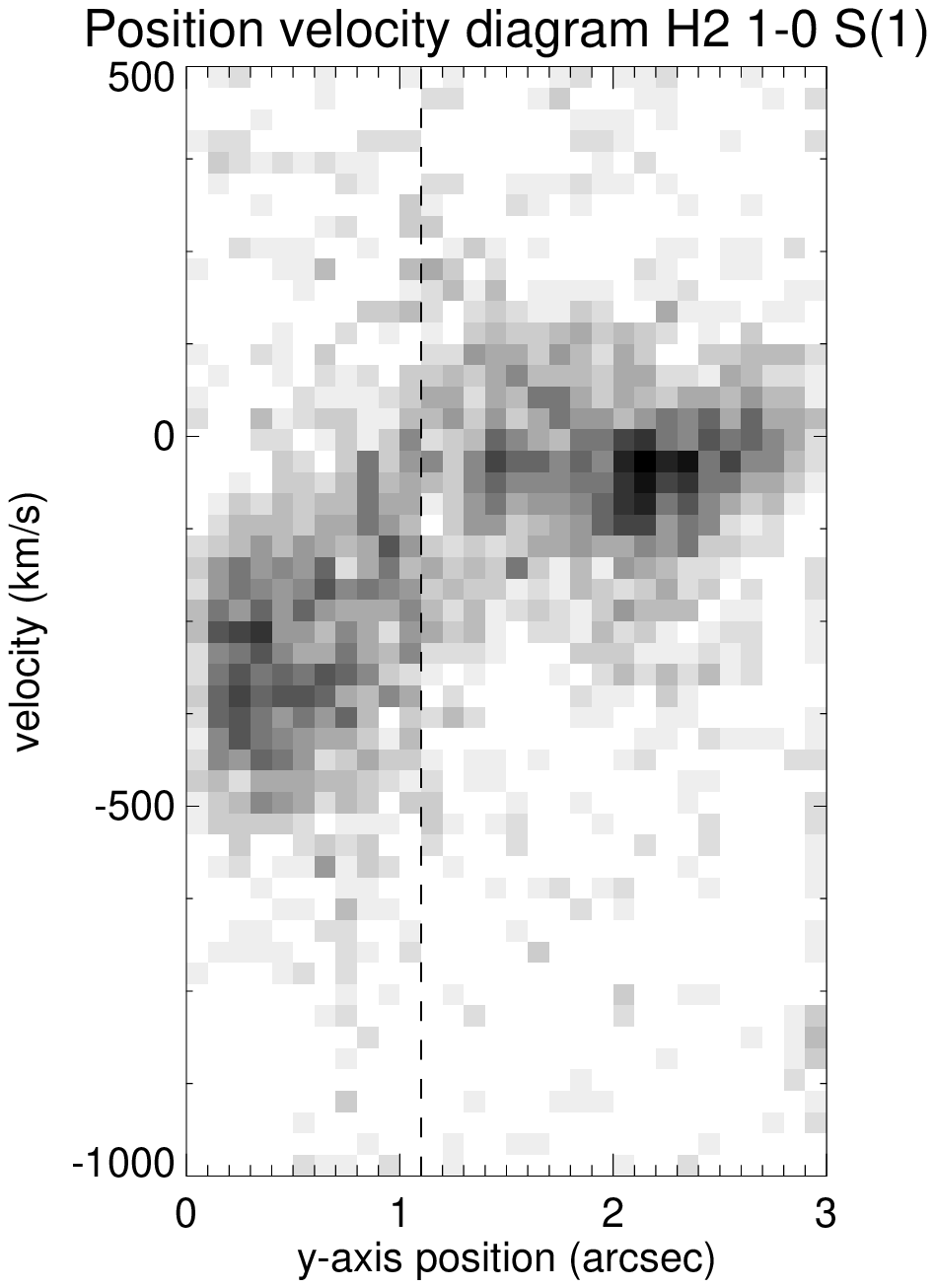}
\caption{Upper: a 1-d cut in the velocity field in a 0.1\arcsec-wide pseudo-slit along the major axis of the \H2~emission indicated in Fig.~\ref{fig:2AH2}. Lower: Position velocity diagram for \H2~v=1-0~S(1) along the same axis (the dashed line indicates the position of the nucleus).}
\label{fig:2AH2kinemcut}
\end{centering}
\end{figure}

The \H2~kinematics exhibit an ordered velocity gradient of more than 300\kmps~over the NIFS field of view. The line width mostly lies in the 
range 200--400\kmps~FWHM, but rises to a peak of 800\kmps~FWHM in the nuclear spaxel and a ridge of elevated velocity dispersion 
(400--600\kmps~FWHM) runs from there through the middle of the south-eastern clump. A second ridge of elevated FWHM lies in the lower 
right of Fig.~\ref{fig:2AH2}, but lies near the edge of the detected emission and is likely to be of lower statistical significance, 
although hints of a similar feature are apparent in the \Ha~observations of Hatch et al.~(2007). In Fig.~\ref{fig:2AH2kinemcut} we show 
an \H2~rotation curve and a position-velocity diagram along the major axis of the \H2~emission. 
 
\subsection{Relationship to \Ha~and CO emission}
There is a close morphological and kinematic correspondence between this \H2~emission and the double-peaked bar of \Ha~emission seen in the IFS data of Hatch et al.~(2007) at lower spatial resolution. Similarly, there is also close agreement between the global \H2~kinematics and
those of the CO emission: in single-dish observations of CO(2-1) emission Edge~(2001) measured a linewidth of 390\kmps~FWHM, offset by 175\kmps~from an 
assumed zeropoint redshift of z=0.0338, which translates to --145\kmps~in our assumed frame (z=0.0349) i.e. the mid-point of the rotation curve in Fig.~\ref{fig:2AH2kinemcut}. Interferometry by Edge \& Frayer~(2003) demonstrated that the CO(1-0) emission arises from both the BCG and the
companion galaxy, with the bulk of the emission residing in the former -- a mass of $1.2 \times 10^{9}$\Msun~of cool molecular gas on a spatial scale $<3.5$\kpc. Given the close spatio-kinematic correspondence, it is reasonable to assume that the \Ha, \H2~and CO emission are all tracing the same gas system.

\begin{figure}
\begin{centering}
\includegraphics[width=6.5cm,angle=0]{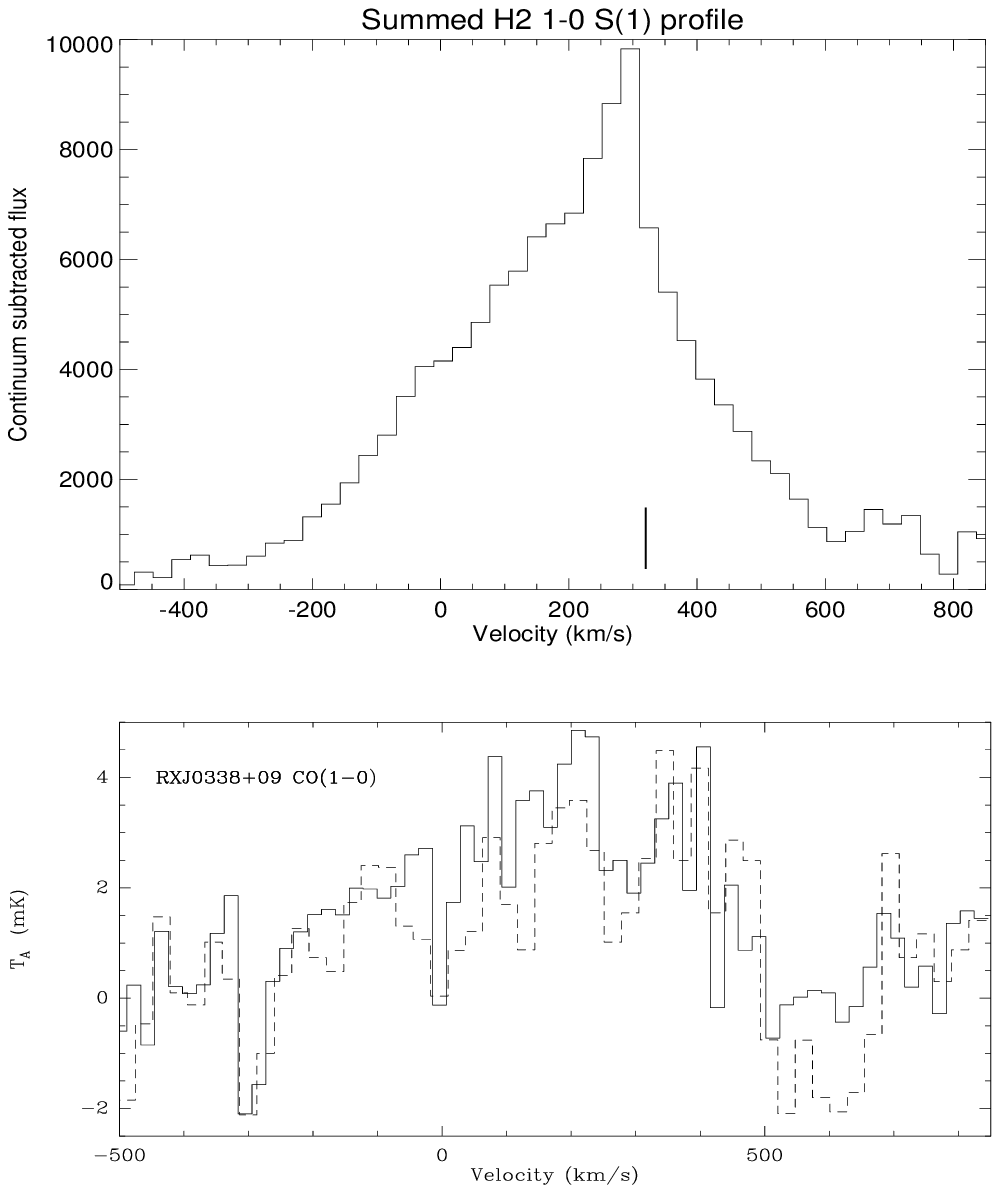}
\caption{Comparison of the integrated continuum-subtracted \H2~v=1-0~S(1) profile within the NIFS field of view with the single dish CO(1--0) profile from Edge (2001). The zeropoint of the velocity system is z=0.0338 to match that used by Edge 
(2001); the position of the z=0.0349 zeropoint used elsewhere in the present paper is indicated with the vertical dash in the upper panel.}
\label{fig:2AH2COcomp}
\end{centering}
\end{figure}

\begin{figure*}
\begin{centering}
\includegraphics[width=0.7\textwidth,angle=0]{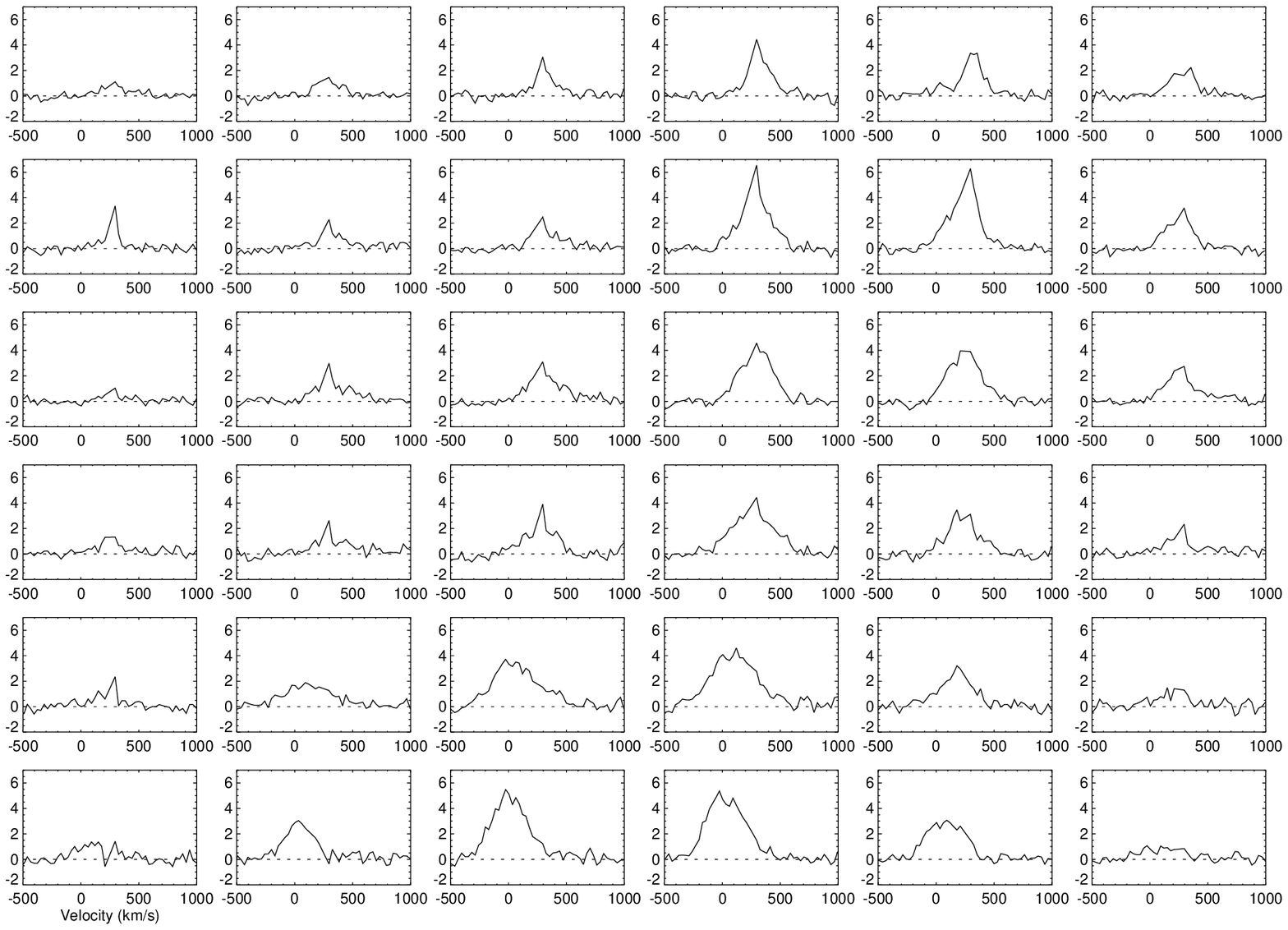}
\caption{\H2~v=1-0~S(1) emission profiles in 0.5\arcsec~cells arranged according to position in the NIFS field of view. A 
flat continuum baseline has been subtracted from each spectrum. The velocity zeropoint used here is z=0.0338 to match Fig.~\ref{fig:2AH2COcomp}.}
\label{fig:2AH2profs}
\end{centering}
\end{figure*}

In Fig.~\ref{fig:2AH2COcomp} we compare the \H2~v=1-0~S(1) line profile integrated over the NIFS aperture
with the single dish CO(1--0) profile from Edge~(2001). Whilst there are some differences in detail -- such as the strong narrow \H2~feature from the north-western clump at +300\kmps -- the two lines span the same overall range in velocity, although proportionally more CO emission arises at larger velocities. The latter may originate from diffuse emission within the BCG, possibly in the form of a jet-driven outflow, as discussed further in section 4.2. Indeed, the spatially-unresolved, compact CO(1--0) emission found by Edge \& Frayer~(2003), which is likely to correspond directly to the \H2~we see with NIFS, accounts for only 60~per cent of the integrated single-dish line intensity measured by Edge~(2001) within the 21.6\arcsec~IRAM beam. There may, however, be some contribution to the single-dish flux from the BCG's companion galaxy. Postage-stamp plots of the \H2~line profiles on a grid of 0.5\arcsec~cells arranged according to position in the NIFS field are shown in Fig.~\ref{fig:2AH2profs}.

\subsection{Excitation of the \H2~and the dearth of circumnuclear star formation}

\begin{figure}
\begin{centering}
\includegraphics[width=6.5cm,angle=0]{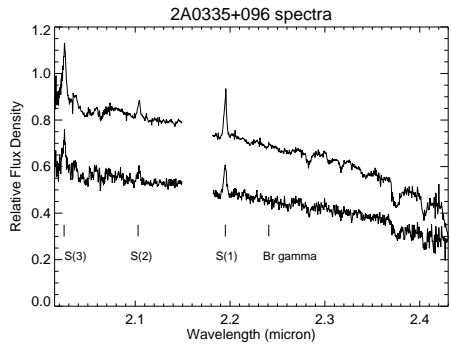}
\caption{Spectra of 2A0335+096 integrated over the NIFS aperture (upper) and in the nucleus (lower; extracted from the box show in Fig.~\ref{fig:2AH2}); the latter is shown scaled up by a factor of 50 and both spectra were boxcar smoothed by 2 pixels in dispersion. The prominent \H2~v=1-0~S(1), S(2) and 
S(3) emission lines are indicated, as is the expected position of the undetected \Bg~emission line. The portions of the spectra around 2.166\micron~were excised due to the presence of \Bg~emission in the telluric standard.}
\label{fig:2AH2spec}
\end{centering}
\end{figure}

Wilman et al.~(2009) demonstrated that the \H2/\Pa~line ratio serves as a good discriminant between \H2~emission excited by star formation (for which \H2~v=1-0~S(3)/\Pa~$\sim 0.2$) and the non-photoionization processes advocated by Ferland et al.~(2009) (for which \H2~v=1-0~S(3)/\Pa~$\sim 1$). In the long-slit spectrum of Edge et al.~(2002) we measured  \H2~v=1-0~S(3)/\Pa=0.79. \Pa~lies just blueward of the spectral coverage of these NIFS data, so we focus instead on \Bg~at 2.1661\micron~(rest) which is within the spectral range. The line is not detected anywhere in the field of view, with a $3\sigma$ upper limit of \Bg/\H2~v=1-0~S(1) $<0.10$ for the integrated spectrum. This limit is consistent with the values of \Bg/\H2~v=1-0~S(1)=0.039 and 0.078 for the `extra heat' and `cosmic ray' cases, respectively, calculated by Ferland et al.~(2009). Fig.~\ref{fig:2AH2spec} shows spectra within the NIFS $3 \times 3$~arcsec aperture and for the $0.2 \times 0.3$~arcsec region of nuclear emission indicated in Fig.~\ref{fig:2AH2}. There is no discernible 
difference in the emission line ratios of the two spectra, suggesting that the excitation source is spatially distributed rather than nucleated. When
combined with the flux ratio of \H2~v=1-0~S(1)/\Pa=0.53~from Edge et al.~(2002) we infer \Bg/\Pa$<0.053$. This is significantly below the case B 
recombination value of 0.083, but closer to the values of 0.058 and 0.064 expected for the `extra heat' and `cosmic ray' filament excitation 
models.

The expected \Bg/\H2~v=1-0~S(1) ratio for star-forming regions is typically much higher. For example, in 
Moorwood \& Oliva~(1988), 7 of out 8 galactic nuclei optically-classified as pure HII systems have \Bg/\H2~v=1-0~S(1)
$ \geq 3$; in composite and Seyfert nuclei, the ratio is of order unity. Our limit is well below this range implying 
that the excitation of \H2~emission in 2A 0335+096 cannot be dominated by star formation. In any case, the star 
formation rate inferred from the mid-infrared emission in 2A 0335+096 by O'Dea et al.~(2008) is just 2.1\Msunpyr~for the entire galaxy, 
compared to $\sim 15$\Msunpyr for the systems in the Wilman et al.~(2009) sample. The $3\sigma$ upper limit on the equivalent width
of \Bg~is 0.8\AA. With reference to the evolutionary models shown in Davies et al.~(2006) (their Fig.~8) the latter would imply 
an absence of star formation for at least the past 10\Myr, if considered as a monolithic stellar population. The measured equivalent width of the CO(2,0) bandhead at 2.2935\micron~(rest) is 9.9\AA; although this figure is below the 12\AA~expected for most stellar populations more than 10\Myr~old (Fig.~6 of Davies et al.~2006), it is likely to lie within the bounds of the systematic uncertainties arising from, for example, the form of the initial mass function in these environments. Modest dilution from non-stellar sources of continuum emission remains another possibility. 

Although we do not detect \Bg~emission within the NIFS aperture, \Ha~emission is present in this region in the IFU observations of Hatch et al.~(2007) and
in the long-slit spectrum of Donahue et al.~(2007). From these works we estimate an \Ha~flux within the $3 \times 3$~arcsec NIFS aperture of
$5 \times 10^{-15}$\ergpcmsqps, equating to an \Ha~luminosity of $1.4 \times 10^{40}$\ergps. Using the Kennicutt~(1998) \Ha--star formation rate 
relation, this would translate to a star formation rate of 0.11\Msunpyr~if all the \Ha~were due to star formation. Corrections for Galactic and
intrinsic dust extinction (each approximately 1 magnitude; Donahue et al.~2007) would boost the estimates by a factor of 6.  As noted by Donahue et 
al.~(2007), the extinction-corrected star formation rate for the entire galaxy inferred from the \Ha~emission is 15--20 \Msunpyr~and that inferred from 
the excess UV emission is of a similar order of magnitude. These estimates would be compatible with the mid-infrared estimate of 2.1\Msunpyr~(O'Dea et al.~2008) if only 10 per cent of the \Ha~emission were due to star formation. For comparison, a fit by Donahue et al.~(2011) to the mid-infrared emission features with an HII region photodissociation region model yielded a star formation rate of 0.7\Msunpyr. 

The low level of ongoing circumnuclear star formation corroborates the findings of Romanishin \& Hintzen~(1988). Their analysis of the B--I colour gradient of the BCG showed that star formation is occurring at radii 4--30\kpc, as evidenced by a plateau in B--I bluer than in non-cooling flow BCGs; the colour reddens at smaller radii, suggesting a decline in the star formation rate or dust extinction. The latter explanation is ruled out by our observations since the extinction in the K-band corresponding to the observed reddening ($\Delta$(B--I) = 0.15) would be completely negligible. We conclude that star formation does not make a significant contribution to the excitation of the \H2~gas within the NIFS field-of-view. In conjunction with the size contraint ($<3.5$\kpc) on the CO emission reservoir from Edge \& Frayer~(2003), this implies that the star formation is spatially offset from the bulk of the molecular gas. We will now address the implications of this finding.

\section{DISCUSSION}

\subsection{Accretion from the \Ha~emission stream and cold feedback}

Based on the evidence presented in section 3, the observed \H2~emission appears to consist of a rotating structure. This could 
take the form of a flared disk or a ring of \H2, or the \H2~peaks could correspond to gas on highly-elliptical orbits which 
spends most of its time at large radii. Given its general alignment with the 14\kpc~stream of \Ha~emission~discussed in section 1, 
it is reasonable to infer that the \H2~has been accreted from this \Ha~stream and may provide a fuel source for the AGN. Indeed, the rotational 
axis of the \H2~aligns with the position angle of the present lobes of radio emission, at least in projection. We now discuss this interpretation further in the context of the cold feedback model. 

Donahue et al.~(2007) argued that the BCG and its nearby companion are interacting and probably gravitationally 
bound to one another. This does not, however, imply that the observed \Ha~stream was stripped from the companion. 
For the case of A1664, Kirkpatrick et al.~(2009a) argued that the stripping of gas from infalling companion 
galaxies was an unlikely origin for the cool gas seen in the cluster core and this generic argument also applies to 2A 0335+096.
Under a stripping scenario, it would in any case be hard to understand why there should be a good spatial correspondence 
between the \Ha~accretion stream and the coolest X-ray gas (Sanders et al.~2009). We speculate that there are at least three 
possible explanations for the association of the \Ha~stream and the companion galaxy. Firstly, the disturbance created by the 
motion of the companion could generate non-linear overdensities in the ICM (of a factor 2 or more) which can readily cool, as 
proposed in the cold feedback model of Pizzolato \& Soker~(2005). Secondly, small amounts of dense molecular gas and dust 
stripped from the companion may offer seeding sites onto which the surrounding ICM can condense and cool, as argued by Sparks 
et al.~(2004) in the case of M87. Thirdly, the companion could have merely disturbed the pre-existing cool gas reservoir of 
the BCG, as argued by Wilman et al.~(2006) for several other BCGs. Alternatively, the companion could be just a chance projection.
There may be some parallels with the radio galaxies 3C~31 and 3C~264 which both exhibit double-horned CO profiles from $\sim 10^{9}$\Msun~of molecular hydrogen, possibly acquired through cannibalism with gas-rich galaxies over the past few tens of \Myr~(Lim et al.~2000).

Once within the BCG, the accretion stream will interact with the local interstellar medium. Depending on the
relative velocity of impact, this collision may generate shocks and disturbed kinematics. The locally elevated \H2~velocity
dispersion in the south-eastern \H2~peak and nearby dust feature (Fig.~\ref{fig:2AHST}) may be signatures of just such an 
interaction. Hydrodynamic interaction may lead to a loss in angular momentum of the accreted material, driving gas into the 
circumnuclear regions as a potential fuel source for the AGN. Mutual collisions among the \H2~cloud ensemble in its dense inner 
regions provide a further mechanism for angular momentum loss, as argued by Pizzolato \& Soker~(2010).

Whilst the smooth rotation curve of Fig.~\ref{fig:2AH2kinemcut} suggests that the \H2~is rotating in the potential of 
the BCG, the existence of two distinct \H2~peaks could suggest that the \H2~clouds are on highly elliptical orbits, with 
the peaks corresponding to their apastrons (see e.g. Fig.~9 of Pizzolato \& Soker~2010). The distortions of the \H2~isovelocity 
contours in Fig.~\ref{fig:2AH2} point to departures from simple circular motions, possibly in the form of disk warping and
elliptical orbits. We note a resemblance to the twin peaks of molecular gas which are commonly observed within the central kiloparsec~of 
barred and starburst galaxies (e.g. Kenney et al.~1992; Jogee et al.~2005). Such peaks result from gas being driven into 
the inner Lindblad resonance (ILR) of the barred potential, which is associated with the transition from the 
highly elongated family of x$_{\rm{1}}$ loop orbits to the smaller and more circular x$_{\rm{2}}$ orbit family. 
The x$_{\rm{1}}$ orbits are aligned with the major axis of the bar, whose leading edges typically exhibit 
radial dust-lanes along which gas flows into the nuclear regions. Whilst there is no obvious evidence for a bar in the BCG, 
the presence of the nearby companion galaxy, or triaxility intrinsic to the BCG itself, may induce a non-axisymmetric potential 
with analogous dynamical consequences for the accreted gas. We are, however, unable to infer whether the observed \H2~peaks 
correspond to gas flowing along the major axis of a bar in the early stages of inflow, or whether the material has already been 
funelled down to the ILR, and do not pursue this interpretation any further.

\subsection{Accretion timescale and outburst energetics}
Given that the accreted \H2~appears to be a viable fuel source for the AGN, we now discuss the associated timescales and energetics.
The current outburst of the radio source gave rise to the observed 8.4\kpc-long lobes in the 5~GHz image of Sanders et al.~(2009). 
The estimated age of this radio emission is 25-50\Myr~(Donahue et al.~2007). If on circular orbits, the orbital timescale for 
the rotating \H2~structure is 25\Myr~for a radius of 600\pc~and a velocity of 150\kmps~(Fig.~\ref{fig:2AH2kinemcut}), neglecting
projection effects. As Sanders et al.~(2009) discussed, the enthalpy of the observed cavities in the cluster core 
($5 \times 10^{59}$\erg) is sufficient to offset the luminosity within the 122\kpc~X-ray cooling radius for a further 
$\sim 50$\Myr, i.e. for a few dynamical times of the \H2. On a similar timescale, we would expect that the central 
supermassive black hole would be able to accrete some of the \H2~in order to sustain the feedback.

The energetics of the outburst arising from accretion from the \H2~reservoir are, of course, dependent on how much of the gas
can be accreted by the black hole and how much is instead channelled into star formation. To maintain strict conformity 
with the black hole:bulge mass scaling relation of Magorrian et al.~(1998), 700\Msun~of star formation would be expected for every 
1\Msun~accreted by the black hole.  Applied to 2A 0335+096, accretion of 1 part in 700 of the $1.2 \times 10^{9}$\Msun~\H2~reservoir at 
the canonical efficiency of $\epsilon=0.1$ would produce an outburst of $3 \times 10^{59}$\erg -- i.e. comparable in size to the
present cavity system measured by Sanders et al.~(2009) -- which could offset cooling for $\sim 30$\Myr. Whilst the rates of star 
formation and black hole growth for a repesentative sample of clusters are in line with such expectations statistically, the sizeable 
scatter shows that they are not growing in precise lockstep in any given system (Rafferty et al.~2006). 

Alternative possibilities stem from the low levels of star formation in the circumnuclear region, which may allow a 
larger fraction of the molecular gas to be accreted. As noted in section 3.3, the current star formation rate within the NIFS aperture, which 
encompasses most if not all of the $1.2 \times 10^{9}$\Msun~molecular gas reservoir, appears to be no more than 1\Msunpyr. At this rate, it would take 
at least 1\Gyr~for the gas to be consumed. The accretion of just $\sim 5$~per cent of the gas onto a moderately-spinning black hole would suffice 
to offset cooling in the cluster core over the same period. This would imply that the AGN would be essentially on for $\sim 1$\Gyr, either continuously 
or in a series of mini-outbursts. The viability of this scenario hinges firstly on whether this current low level of star formation can be sustained, 
and secondly on whether the gas can be prevented from accreting much more quickly, given that the orbital timescale of the observed \H2~is just 25\Myr. 
Indeed, from a comparison of the relative \H2~line strengths in the nuclear and integrated spectra (Fig.~\ref{fig:2AH2spec}), we infer that $\sim 2$~per cent of the molecular gas ($2.5 \times 10^{7}$\Msun) is already within 100\pc~of the nucleus. These two issues go to the heart of understanding the mechanisms which regulate cold feedback, on which we speculate below.

One explanation could be that the current radio outburst may have initially evacuated gas from the inner regions or otherwise suppressed 
star formation; the resolved lateral extent of the current 5~GHz radio source (i.e. perpendicular to the inferred jet direction) in 
Sanders et al.~(2009) is $\pm 5$\arcsec~(3.5\kpc). The molecular gas could be supported by the associated magnetic 
field and thus prevented from collapsing to form stars, as argued by Fabian et al.~(2008) and Ferland et al.~(2009) for the \Ha~filaments in 
NGC 1275 and the excitation of their line emission. It is also possible that the current star formation rate is unrepresentative 
of the average value on Gyr timescales; the star formation rate could increase and consume the gas more quickly in the future. 

Magnetic support of the gas, if sufficiently strong, could hinder the accretion onto the black hole. Outflows driven by the radio source provide a further means to disrupt its fuel supply. In section 3.2, we noted that some of the diffuse CO emission within the BCG is found at relatively high velocity, suggestive of an outflow. Less than 1 per cent of the current outburst enthalpy would
suffice to accelerate $10^{9}$\Msun~of molecular gas to a velocity of 600\kmps, so the energetics are feasible and similar to the 
fraction of the outburst energy required to drive the metal outflows in clusters (Kirkpatrick et al. 2009b, 2011). If such mechanisms are, 
however, not effective in regulating the accretion and the bulk of the \H2~is instead accreted rapidly by the black hole over several orbital 
timescales, this could generate an outburst exceeding $10^{62}$\erg~(for $\epsilon=0.1$). Such an outburst would be comparable with most powerful 
currently known, in MS0735.6+7421 (McNamara et al.~2005) and Hydra A (Wise et al.~2007). Further simulations of cold feedback 
including the effects of magnetic fields and outflows on the molecular gas are required to discriminate between such possibilities.
 
Not withstanding these uncertainties, the new observations of 2A 0335+096 presented here, revealing an ample supply 
of cool molecular gas available to fuel the radio source on the required timescale of 25-100\Myr~and offset ICM cooling, further 
bolster the case for cold AGN feedback.

\section*{ACKNOWLEDGMENTS}
Based on observations obtained at the Gemini Observatory, which is operated by the 
Association of Universities for Research in Astronomy, Inc., under a cooperative agreement 
with the NSF on behalf of the Gemini partnership: the National Science Foundation (United 
States), the Science and Technology Facilities Council (United Kingdom), the 
National Research Council (Canada), CONICYT (Chile), the Australian Research Council (Australia), 
Minist\'{e}rio da Ci\^{e}ncia e Tecnologia (Brazil) 
and Ministerio de Ciencia, Tecnolog\'{i}a e Innovaci\'{o}n Productiva (Argentina). Program ID 
GN-2009B-Q-69. BRM acknowledges generous financial support from the Canadian Space Agency Space 
Science Enhancement Program, and from the Natural Sciences and Engineering Research Council of 
Canada's Discovery Grant program.

{}

\end{document}